\documentclass[12pt]{article}
\usepackage{graphics}
\usepackage{amsfonts}
\usepackage{amsmath}
\usepackage{graphicx,epsfig,float}
\usepackage{bm}

\def\R{\mathbb{R}}
\def\N{\mathbb{N}}
\def\I{1\!\!1}
\newtheorem{theorem}{Theorem}

\newtheorem{definition}[theorem]{Definition}

\newtheorem{proposition}[theorem]{Proposition}

\begin{document}

\title{Chaos Decomposition and Gap Renormalization of
Brownian Self-Intersection Local Times}
\author{\textbf{Jinky Bornales} \\
{\small Physics Department, MSU-IIT, Iligan City, The Philippines}\\
{\small jinky.bornales@g.msuiit.edu.ph}
\and \textbf{Maria Jo\~{a}o Oliveira} \\
{\small Universidade Aberta, P 1269-001 Lisbon, Portugal}\\
{\small CMAF, University of Lisbon, P 1649-003 Lisbon, Portugal}\\
{\small oliveira@cii.fc.ul.pt, mjoliveira@ciencias.ulisboa.pt (New)} \and \textbf{Ludwig Streit} \\
{\small Forschungszentrum BiBoS, Bielefeld University, D 33501 Bielefeld,
Germany}\\
{\small CCM, University of Madeira, P 9000-390 Funchal, Portugal}\\
{\small streit@physik.uni-bielefeld.de}}
\date{}
\maketitle

\begin{abstract}
We study the chaos decomposition of self-intersection local times and their
regularization, with a particular view towards Varadhan's renormalization
for the planar Edwards model.
\end{abstract}


\smallskip


\section{Introduction}

The self-intersection local time of $d$-dimensional Brownian motion,
informally, is given as 
\begin{equation}
L=\int_{0}^{T}dt_2\int_{0}^{t_2}dt_1\,\delta \left(\mathbf{B}(t_2)-\mathbf{B}(t_1)\right) .  \label{L}
\end{equation}%
We shall see that, while "reasonably well defined" for $d=1$, these local
times become more and more singular as the dimension $d$ increases.
Intersections have thus been the object of extensive study by authors
such as Dvoretzky, Erd\"os, Kakutani \cite{Dv}, \cite{Dv1}, \cite{Dv2}, 
Varadhan \cite{v}, Westwater \cite{Westwater}, \cite{Westwater2}, \cite{Westwater3},
Le Gall \cite{LeGall}, \cite{LeGall2}, Rosen \cite{R1}, \cite{R2}, \cite{R3}, Dynkin
\cite{Dy1}, \cite{Dy2}, \cite{Dynkin}, Watanabe \cite{Wat}, 
Yor \cite{Yor1}, \cite{Yor2}, Imkeller et al.~\cite{Imkeller}, Albeverio
et al.~\cite{AOS}, \cite{AlbHu}. For fractional Brownian motion there are papers 
e.g.~by Rosen \cite{R4}, Hu \& Nualart \cite{Hu1}, Grothaus et al.~\cite{GOSS}.

Apart from its intrinsic mathematical interest the self-intersection local
time has played a role in constructive quantum field theory, and is a
standard model in polymer physics for the self-repulsion ("excluded volume
effect") of chain polymers in solvents \cite{Schaefer}.

Replacement of the Dirac delta function in (\ref{L}) by a Gaussian 
\begin{equation*}
\delta _{\varepsilon }(x):=\frac{1}{(2\pi \varepsilon )^{d/2}}e^{-\frac{%
|x|^{2}}{2\varepsilon }},\quad \varepsilon >0,
\end{equation*}
leads to regularized local times
\begin{equation*}
L_{\varepsilon }:=\int_{0}^{T}dt_2\ \int_{0}^{t_2}dt_1\,\delta _{\varepsilon
}(\mathbf{B}(t_2)-\mathbf{B}(t_1))   
\end{equation*}
and for $d=1$ one can show $L^{2}$ convergence w.r.t.~white noise or Wiener
measure space. But already for $d=2$ this fails since the expectation of 
$L_{\varepsilon }$ will diverge in the limit, asymptotically
\begin{equation*}
\mathbb{E}(L_{\varepsilon })\approx -\frac{T}{2\pi }\ln \varepsilon.
\end{equation*}
In this case it is sufficient to subtract the expectation, i.e.~the centered
regularized local time does have a well-defined $L^2$ limit:
\begin{equation*}
L_{\varepsilon ,c}:=L_{\varepsilon }-\mathbb{E}(L_{\varepsilon })\rightarrow
L_{c}.
\end{equation*}
Apart from the Gaussian regularization above, others have been considered to
remove the singularity at $t_1=t_2$ in the integral (\ref{L}). The "staircase
regularization" avoids the line $t_1=t_2$ as in see e.g.~Bolthausen 
\cite{Bolthausen} (Fig.~1). 
\begin{center}
\includegraphics[scale=1.0]{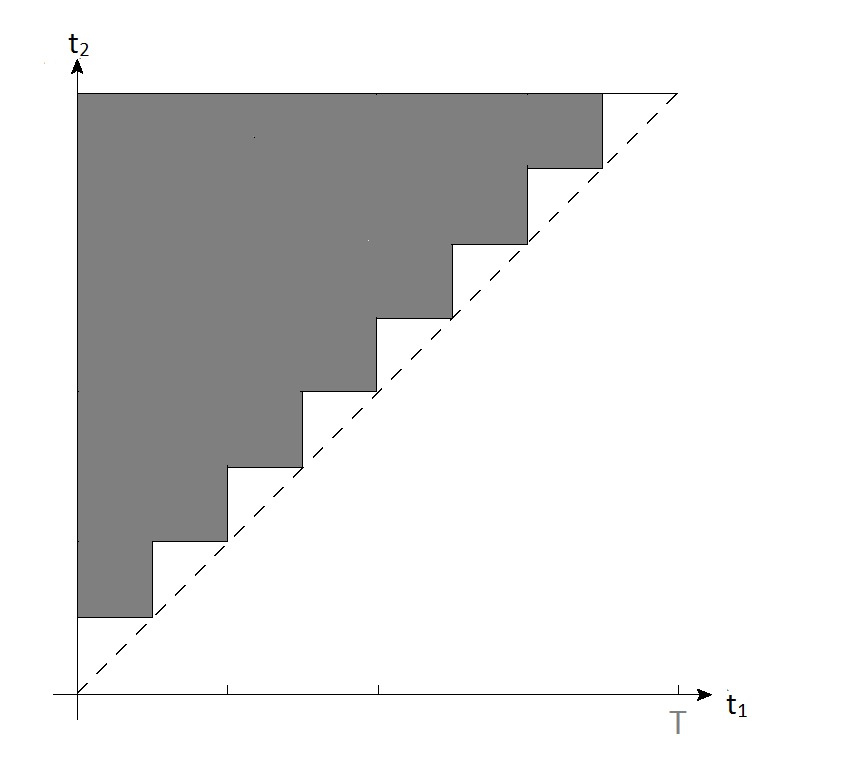}
\end{center}
{\small {\bf Fig.~1}: Domain of integration for the staircase-regularized local time.}

\bigskip

\noindent
The widely used "gap regularization" does the same
by omitting the strip $t_2-t_1<\Lambda$ in the integral. In the modelling of
chain polymers the gap size $\Lambda$ will be a "microscopic" quantity,
i.e.~of the order of the inter-monomer distance, more precisely the "Kuhn"
or "persistence" length. It plays an important role in renormalization group
calculations \cite{Schaefer}: critical parameters are obtained from the postulate that
macroscopic quantities do not depend on microscopic length scales.

\section{Tools from White Noise Analysis \cite{hkps}}

Based on a $d$-tuple of independent Gaussian white noises 
$\bm{\omega}=(\omega _{1},...,\omega _{d})$ one 
defines a $d$-dimensional Brownian motion $\mathbf{B}$ through
\begin{equation*}
\mathbf{B}(t)\equiv \langle\bm{\omega},\I_{[0,t]}\rangle=\int_{0}^{t}ds\,
\bm{\omega }(s).
\end{equation*}

We shall use a multi-index notation
\begin{equation*}
\mathbf{n}=(n_{1},\ldots ,n_{d}),\;\;n=\sum_{i=1}^{d}n_{i},\;\;\mathbf{n}%
!=\prod_{i=1}^{d}n_{i}!
\end{equation*}%
and for $d$-tuples of Schwartz test functions $\mathbf{f}=(f_{1},\ldots,f_{d})\in 
S(\mathbb{R},\mathbb{R}^{d})$,
\begin{equation*}
\langle\mathbf{f},\mathbf{f}\rangle=\sum_{i=1}^{d}\int dt\, f_{i}^{2}(t)
\end{equation*}
\begin{equation*}
\langle F_{\mathbf{n}},\mathbf{f}^{\otimes \mathbf{n}}\rangle=\int d^nt\, 
F_{\mathbf{n}}(t_1,\ldots,t_n) \underset{i=1}{\overset{d}{\otimes }}
f_i^{\otimes n_i}(t_1,\ldots,t_n)
\end{equation*}
and similarly for $\langle:\bm{\omega}^{\otimes \mathbf{n}}:,F_{\mathbf{n}}
\rangle$ where for $d$-tuples of white noise the Wick product $: \cdot :$ 
\cite{hkps} generalizes to 
\begin{equation*}
:\bm{\omega }^{\otimes \mathbf{n}}:
=\underset{i=1}{\overset{d}{\otimes }}:\omega _{i}^{\otimes n_{i}}:
\end{equation*}
The vector valued white noise $\omega$ has the characteristic function 
\begin{equation*}
C(\mathbf{f}):=\mathbb{E}(e^{i\langle\bm{\omega },\mathbf{f}\rangle})=\int_{S^{\ast }(\mathbb{R%
},\mathbb{R}^{d})}d\mu(\bm{\omega }) e^{i\langle\bm{\omega },%
\mathbf{f}\rangle}=e^{-\frac{1}{2}\langle\mathbf{f},\mathbf{f}\rangle},
\end{equation*}
where $\langle\bm{\omega},\mathbf{f}\rangle
=\sum_{i=1}^d\langle\omega_{i}, f_{i}\rangle$ and $f_{i}\in S(\R, \R)$.

Writing 
\begin{equation*}
(L^{2}):= L^{2}(S^{\ast }(\R,\R^{d}), d\mu)
\end{equation*}
there is the It\^o-Segal-Wiener isomorphism with the Fock space of symmetric
square integrable functions: 
\begin{equation*}
(L^{2})\simeq \left(\underset{k=0}{\overset{\infty }{\oplus }}\mathrm{Sym}\, 
L^{2}(\mathbb{R}^{k},k!d^{k}t)\right)^{\otimes d}.  
\end{equation*}
This implies the chaos expansion 
\begin{equation*}
\varphi (\bm{\omega })=\sum_{\mathbf{n}\in\N_0^d}
\langle:\bm{\omega }^{\otimes \mathbf{n}}:,F_{\mathbf{n}}\rangle\textrm{ for  }
\varphi \in (L^{2})  
\end{equation*}
with kernel functions $F_{\mathbf{n}}$ in Fock space. 

Generalized functionals are constructed via a Gel'fand triple 
\begin{equation*}
(S)\subset (L^{2})\subset (S)^{\ast }.
\end{equation*}

The generalized functionals in ${(S)}^{\ast }$ are conveniently
characterized by their action on exponentials. In particular we use the 
\begin{equation*}
:\exp (\langle\bm{\omega},\mathbf{f}\rangle):\,= C(\mathbf{f})
\exp(\langle\bm{\omega},\mathbf{f}\rangle)\in (S)  
\end{equation*}
to make the

\begin{definition} The transformation defined for all test functions 
$\mathbf{f}\in S(\R,\R^d)$ via the bilinear dual product on 
${(S)}^{*}\times (S)$ by 
\begin{equation*}  
(S\Phi)(\mathbf{f})=
\langle\!\langle\Phi,:\exp (\langle\cdot,\mathbf{f}\rangle):\rangle\!\rangle
\end{equation*}
is called the S-transform of $\Phi\in {(S)}^{*}$. 
\end{definition}

The multilinear expansion of $S(\Phi)$
\begin{equation*}
(S\Phi)(\mathbf{f})=\sum_{\mathbf{n}\in\N_0^d}\langle
\bm{\varphi}_{\mathbf{n}},\mathbf{f}^{\otimes \mathbf{n}}\rangle  
\end{equation*}
extends the chaos expansion to $\Phi \in (S)^{\ast }$, with distribution
valued kernels $\varphi_{\mathbf{n}}$, such that 
\begin{equation*}
\langle\!\langle\Phi,F\rangle\!\rangle=
\sum_{\mathbf{n}\in\N_0^d}\mathbf{n}!\langle\bm{\varphi}_{\mathbf{n}},
F_{\mathbf{n}}\rangle.
\end{equation*}

\begin{definition}
We shall indicate the projection onto chaos of order $n\geq k$ by a
superscript $(k)$:
\begin{equation*}
\langle\!\langle\Phi ^{(k)},F\rangle\!\rangle=
\sum_{\mathbf{n}: n\geq k}\mathbf{n}!\langle\bm{\varphi }_{\mathbf{n}},
F_{\mathbf{n}}\rangle.
\end{equation*}
\end{definition}

\begin{proposition}
\cite{fhsw}
\begin{equation*}
\delta ^{(2N)}(\mathbf{B}(t_{2})-\mathbf{B}(t_{1}))\in (S)^{\ast }
\end{equation*}
with even kernel functions 
\begin{equation*}
\bm{\psi }_{2\mathbf{n}}(u_{1},\ldots
,u_{2n};t_{1},t_{2})=\frac{1}{\mathbf{n}!}(2\pi )^{-d/2}
\left(\frac{1}{|t_{2}-t_{1}|}\right)^{\frac{d}{2}+n}
\left(-\frac{1}{2}\right)^{n}
\prod_{k=1}^{2n}\I_{[t_{1},t_{2}]}(u_{k}).  
\end{equation*}
All the kernel functions with odd indices vanish.
\end{proposition}

Setting 
\begin{eqnarray*}
v &:=&\max (u_{1},\ldots ,u_{2n}) \\
u &:=&\min (u_{1},\ldots ,u_{2n})
\end{eqnarray*}%
one computes \cite{fhsw} the kernel functions of the 
truncated local time $L^{(2N)}$ for $2N>d-2$ by integration over 
$0<t_{1}<t_{2}<T$:
\begin{eqnarray*}
\varphi _{2\mathbf{n}}(u_{1},\ldots ,u_{2n})&=&\frac{(2\pi
)^{-d/2}}{\mathbf{n}!}\left(-\frac{1}{2}\right)
^{n}\int_{0}^{T}dt_{2}\int_{0}^{t_{2}}dt_{1}
\,(t_{2}-t_{1})^{-n-d/2}\I_{[t_{1},t_{2}]}^{\otimes 2n}(u_{1},\ldots ,u_{2n})\\ 
&=&(-1)^{n}\left((n+\frac{d}{2}-1)(n+\frac{d}{2}-2)(2\pi
)^{d/2}\,2^{n}\,\mathbf{n}!\right)^{-1}\cdot \Theta(u)\Theta (T-v)\cdot \\ 
&&\cdot (T^{-n-\frac{d}{2}+2}-v^{-n-\frac{d}{2}+2}-(T-u)^{-n-\frac{d}{2}%
+2}+(v-u)^{-n-\frac{d}{2}+2})%
\end{eqnarray*}
except for $2n=d=2$ where
\begin{equation*}
\varphi _2(u_{1},u_{2})=-\frac{1}{4\pi}\left(\ln v+\ln (T-u)-\ln
(v-u)-\ln T\right)\cdot\Theta(u)\Theta(T-v).
\end{equation*}
The Heaviside function $\Theta$ here is the 
indicator function of the positive half line.

\section{Regularizations}

Replacement of the Dirac delta function by a Gaussian 
\begin{equation*}
\delta _{\varepsilon }(x)=\frac{1}{(2\pi\varepsilon )^{d/2}}e^{-\frac{%
|x|^{2}}{2\varepsilon}},\quad \varepsilon >0,
\end{equation*}%
leads to regularized local times
\begin{equation}
L_{\varepsilon }=\int_{0}^{T}dt_{2}\int_{0}^{t_{2}}dt_{1}\,\delta
_{\varepsilon }(\mathbf{B}(t_{2})-\mathbf{B}(t_{1}))   \label{3Eq2}
\end{equation}
with kernel functions \cite{fhsw}
\begin{eqnarray*}
&&\varphi _{\varepsilon,2\mathbf{n}}(u_{1},\ldots ,u_{2n})\\
&=&\frac{(2\pi )^{-d/2}}{\mathbf{n}!}\left(-\frac{1}{2}\right)
^{n}\int_{0}^{T}dt_{2}\int_{0}^{t_{2}}dt_{1}\,(\varepsilon +\left|
t_{2}-t_{1}\right|)^{-n-d/2}\I_{[t_{1},t_{2}]}^{\otimes 2n}(u_{1},\ldots
,u_{2n}) \\ 
&=& (-1)^{n}\left((n+\frac{d}{2}-1)(n+\frac{d}{2}-2)(2\pi
)^{d/2}\,2^{n}\,\mathbf{n}!\right)^{-1}\cdot \Theta
(u)\Theta (T-v)\cdot \\ 
&&\cdot ((T+\varepsilon)^{-n-\frac{d}{2}+2}-(v+\varepsilon )^{-n-\frac{d}{2}
+2}-(T-u+\varepsilon)^{-n-\frac{d}{2}+2}+(v-u+\varepsilon )^{-n-\frac{d}{2}%
+2}),
\end{eqnarray*}
\begin{eqnarray*}
&&\varphi _{\varepsilon,2}(u_{1},u_{2})\\
&=&-\frac{1}{4\pi}\left( \ln
(v+\varepsilon)+\ln (T-u+\varepsilon)-\ln (v-u+\varepsilon )-\ln
(T+\varepsilon )\right)\cdot\Theta(u)\Theta(T-v).
\end{eqnarray*}

\subsection{Gap Regularization of Kernel Functions}

In renormalization group studies of self-repelling Brownian motion another
regularization is often used, see e.g.~\cite{Schaefer} and the references
there; it suppresses intersection in small time intervals $t_{2}-t_{1}$
between intersections by setting informally
\begin{equation*}
L(\Lambda ):=\underset{t_{2}-t_{1}>\Lambda }{\underset{0<t_{1}<t_{2}<T}
{\int d^{2}t}}\delta (\mathbf{B}(t_{2})-\mathbf{B}(t_{1})).
\end{equation*}

The expectation of $L(\Lambda)$ is equal to
\begin{equation*}
\mathbb{E}\left( L(\Lambda )\right) =\int_{\Lambda
}^{T}dt_{2}\int_{0}^{t_{2}-\Lambda }dt_1\,\bm{\psi }_{0}(t_{1},t_{2})=(2\pi
)^{-d/2}\int_{\Lambda }^{T}dt_{2}\int_{0}^{t_{2}-\Lambda
}dt_{1}\,(t_{2}-t_{1})^{-d/2}.
\end{equation*}
We note that for in particular $d=2$ 
\begin{equation}
\mathbb{E}\left( L(\Lambda )\right) =-\frac{T}{2\pi }\ln \Lambda +O(1).
\label{eqRef}
\end{equation}
Recall \cite{fhsw} that for $\Lambda=0$ the kernel functions $\varphi_{2\mathbf{n}}$ 
of the truncated local time $L^{(2N)}$ are obtained by
integrating the kernel functions $\psi_{2\mathbf{n}}(u_{1},\ldots,
u_{2n};t_{1},t_{2})$ over the rectangle $0<t_{1}<u$ and $v<t_{2}<T$, 
shaded grey in see Fig.~2. For $\Lambda >0$ the integration is further 
restricted to the light domain with $t_{2}-t_{1}>\Lambda$. This restriction 
is non-trivial when $v-u<\Lambda$.
\begin{equation*}
L^{(2N)}-L^{(2N)}(\Lambda )
\end{equation*}
thus has kernel functions $\rho_{2 \mathbf{n}}(u,v)$ with support on
\begin{equation*}
\left\{ 0<u<v<T\right\} \cap \left\{ v-u<\Lambda \right\}.
\end{equation*}
Fig.~2 is pertinent to the case where $\Lambda <v$ and $u<T-\Lambda$. In
this case the kernel functions $\rho _{2 \mathbf{n}}(u,v)$ are obtained by integrating
the $\psi _{2\mathbf{n}}(u_{1},\ldots, u_{2n};t_{1},t_{2})$ with respect to the $t_{i}$
over $v-\Lambda <t_{1}<u$ and $v\ <t_{2}<t_{1}+\Lambda$. Excepting the
kernel function $\rho _{2}(u,v)$ for $d=2$, one finds,
\begin{eqnarray}
\rho _{2\mathbf{n}}(u,v) &=&\int_{v-\Lambda}^{u}dt_{1}\int_{v}^{t_{1}+\Lambda }dt_{2}\,
\psi _{2\mathbf{n}}(u_{1},\ldots, u_{2n};t_{1},t_{2})\label{range} \\
&=&\frac{(2\pi)^{-d/2}}{\mathbf{n}!}\left(-\frac{1}{2}\right)^{n}
\frac{1}{d/2+n-1}\cdot\nonumber \\
&&\cdot\left(\frac{1}{d/2+n-2}\left((v-u)^{-d/2-n+2}-\Lambda^{-d/2-n+2}\right)+
(v-u-\Lambda)\Lambda^{-d/2-n+1}\right)\nonumber
\end{eqnarray}
and for $2n=d=2$ 
\begin{equation}
\rho_{2}(u,v)=\frac{1}{4\pi}\left(\ln (v-u)-\ln \Lambda +\frac{\Lambda
-v+u}{\Lambda }\right).  \label{d}
\end{equation}
\begin{center}
\includegraphics[scale=1.0]{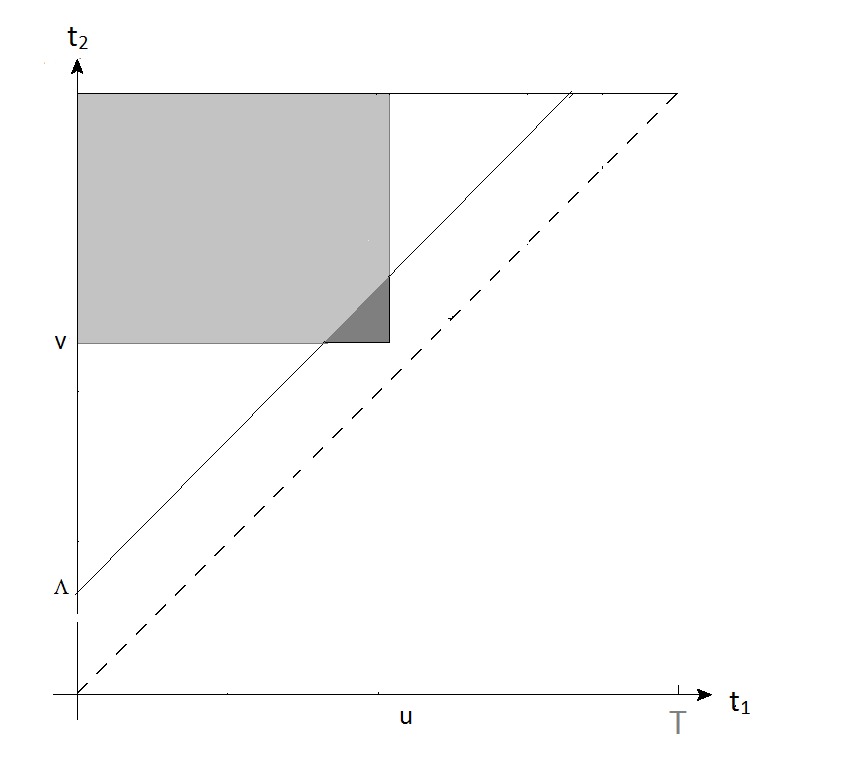}
\end{center}
{\small {\bf Fig.~2}: Domain of integration for kernels of the local time, light grey for the regularized local 
time, dark grey for the subtraction $\rho$ as in \eqref{range}.}

Using $\tau =t_{2}-t_{1}$ we obtain the following estimate
\begin{eqnarray}
\left| \rho _{2\mathbf{n}}(u,v)\right|  &=&\left|
\int_{v-\Lambda}^{u}dt_{1}\int_{v}^{t_{1}+\Lambda }dt_{2}\,
\psi _{2\mathbf{n}}(u_{1},\ldots, u_{2n};t_{1},t_{2})\right|  \label{range2} \\
&=&\frac{1}{2^n\mathbf{n}!}(2\pi)^{-d/2}
\int_{v-\Lambda}^{u}dt_{1}\int_{v-t_{1}}^{\Lambda }d\tau\, 
\left(\frac{1}{\tau }\right)^{\frac{d}{2}+n} \nonumber\\
&\leq &\frac{1}{2^n\mathbf{n}!}(2\pi)^{-d/2}
\frac{1}{d/2+n-1}\int_{v-\Lambda}^{u}dt_{1}\,\left( v-t_{1}\right) ^{-d/2-n+1} \nonumber\\
&\leq &\frac{1}{2^n\mathbf{n}!}(2\pi)^{-d/2}
\frac{1}{d/2+n-1}\frac{1}{d/2+n-2}\left(v-u\right)^{-d/2-n+2}\nonumber
\end{eqnarray}
while for $d=2n=2$ one readily finds from (\ref{d})
\begin{equation}
\left| \rho_2(u,v)\right|  
\leq \frac{1}{4\pi}\left| \ln (v-u)\right| \label{ln}
\end{equation}
when $v-u<\Lambda \ll 1$.

For very small or very large $u, v$, i.e.~$0<u<v<\Lambda$ or $T-\Lambda
<u<v<T$, respectively, the range of integrations in (\ref{range}) and 
(\ref{range2}) for $\psi _{2\mathbf{n}}(u_{1},\ldots, u_{2n};t_{1},t_{2})$
will be $0<t_{1}<u$ and $v<t_{2}<t_{1}+\Lambda$, or $v<t_{2}<T$
and $t_{2}-\Lambda <t_{1}<u$ respectively, see Fig.~3. 
\begin{center}
\includegraphics[scale=1.0]{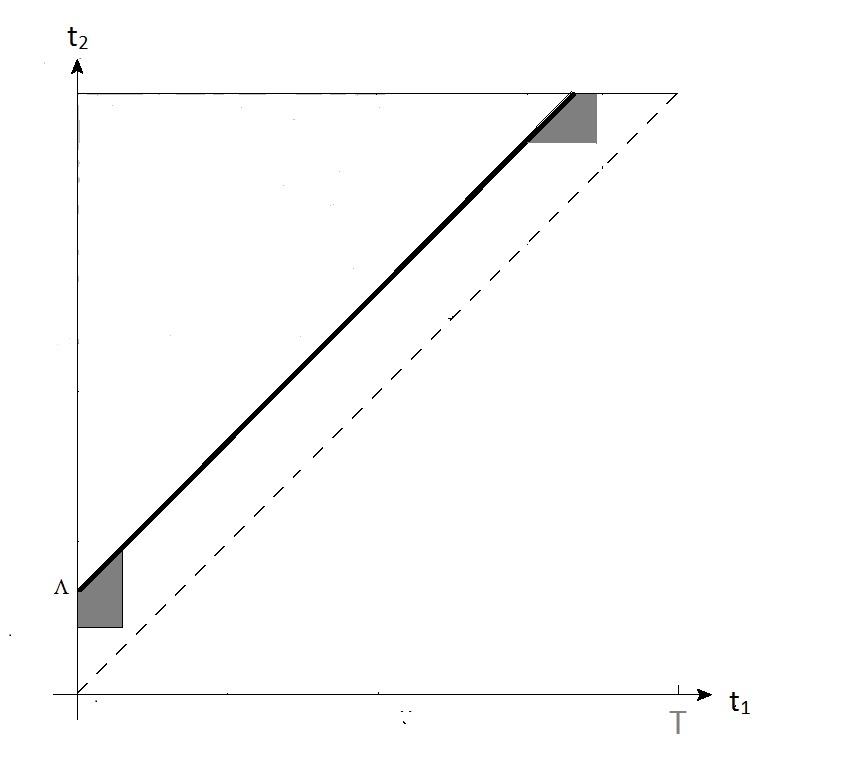}
\end{center}
{\small {\bf Fig.~3}: Modified integration domains for $\rho$ when $u, v$ are close to zero or to $T$ respectively.}

\bigskip

\noindent
Computations as in 
(\ref{range}) are again straightforward, we note here only that the estimate
of (\ref{range2}) is true also in these\ two cases.

\begin{theorem}
For $N>0$ and $0<\Lambda\ll T$ the chaos expansion of the
gap-regularized local time $L^{(2N)}(\Lambda)$ has the kernel functions
\begin{equation}
\varphi_{\Lambda,2\mathbf{n}}(u_{1},\ldots ,u_{2n})=
\varphi _{2\mathbf{n}}(u_{1},\ldots ,u_{2n})-
\Theta(\Lambda -(v-u)) \rho_{2\mathbf{n}}(u_{1},\ldots ,u_{2n}),
\label{jinky}
\end{equation}
and zero otherwise, while 
\begin{equation*}
L^{(2N)}-L^{(2N)}(\Lambda )
\end{equation*}
has the kernel functions $\rho_{2\mathbf{n}}$ for $0<u<v<T$, $n\geq N$.
\end{theorem}

The Heaviside function $\Theta$ exhibits the support property of the 
$\rho_{2\mathbf{n}}$, i.e.~in the gap regularization the kernel functions
of the local time are only modified when all arguments $u_{k}$ are close to
each other.

With these results one can in particular estimate the rate of
convergence for the centered self-intersection local time in $d=2$. Apart
from the term $n=1$ the sum
\begin{equation*}
\left\|L^{(2)}-L^{(2)}(\Lambda)\right\|^2_{(L^2)}
=\sum_{\mathbf{n}: n\geq 1}(2\mathbf{n})!\left\|\rho_{2\mathbf{n}}\right\|^2_{L^2([0,T]^{2n})}
\end{equation*}
can be estimated as follows: 
\begin{eqnarray*}
\sum_{\mathbf{n}: n>1}\left( 2\mathbf{n}\right)!
\left\|\rho _{2\mathbf{n}}\right\|_{L^{2}(\left[ 0,T\right] ^{2n})}^{2}
&\leq&(2\pi )^{-d}\sum_{\mathbf{n}: n> 1}
\frac{\left( 2\mathbf{n}\right) !}{(\mathbf{n}!) ^{2}}
\left(\frac{1}{2}\right)^{2n}\left( \frac{1}{d/2+n-1}\frac{1}{d/2+n-2}%
\right) ^{2} \\
&&\underset{v-u<\Lambda}{\int_{0}^{T}}d^{2n}u_{k}\,
\left(\frac{1}{v-u}\right)^{d+2n-4}. 
\end{eqnarray*}
We can integrate out the $2n-2$ variables $u_{k}$ with $u<u_{k}<v$ that lie
between the smallest and the largest and obtain in this way 
\begin{eqnarray*}
\sum_{\mathbf{n}: n>1}\left( 2\mathbf{n}\right)!
\left\|\rho _{2\mathbf{n}}\right\|_{L^{2}(\left[ 0,T\right] ^{2n})}^{2}
&\leq &(2\pi)^{-2}\sum_{\mathbf{n}: n>1}
\frac{\left( 2\mathbf{n}\right) !}{( \mathbf{n}!)^{2}}
\left(\frac{1}{2}\right)^{2n}\left( \frac{1}{n}\frac{1}{n-1}\right) ^{2} \\
&&\cdot 2n(2n-1)\underset{v-u<\Lambda }{\int_{0}^{T}dv\int_{0}^{v}du}\\
&\leq&\frac{\Lambda T}{2\pi^2}\sum_{\mathbf{n}: n>1}\frac{
\left( 2\mathbf{n}\right) !}{2^{2n}\left(\mathbf{n}!\right)^{2}}
\frac{1}{n}\frac{2n-1}{(n-1)^2}.
\end{eqnarray*}
The series is convergent (Stirling's formula). From \eqref{ln} it is 
straightforward to estimate the remaining term with $n=1$:
\begin{equation*}
\left\|\rho_2\right\|^2_{L^2([0,T]^2)}\leq
\left(\frac{1}{4\pi}\right)^2\int_0^Tdt\int_0^\Lambda
d\tau\,\ln^2\tau=0(T\Lambda \ln ^{2}\Lambda).
\end{equation*}
So we have shown
\begin{theorem}\label{Prop}
For $d=2$
\begin{equation*}
\left\| L^{(2)}-L^{(2)}(\Lambda )\right\|_{\left( L^{2}\right)
}^{2}=0(T\Lambda \ln ^{2}\Lambda)\text{ as }\Lambda \searrow 0.
\end{equation*}
\end{theorem}

A similar improvement of the rate of convergence has been found in the
Gaussian regularization in \cite{Bock}.

\section{Varadhan Renormalization}

The model proposed by Edwards \cite{Edwards} for self-repelling Brownian
motion suppresses self-crossings, modifying the Brownian path (or white
noise) measure by a density function, informally
\begin{equation*}
\varrho =Z^{-1}\ \exp \left( -gL\right)
\end{equation*}
with $g>0$ 
\begin{equation*}
Z=\mathbb{E}\left( \exp \left( -gL\right) \right) .
\end{equation*}

There is no problem for $d=1$ since $L$ is a positive random variable and
hence $\exp \left( -gL\right) <1$. For $d=2$ however we should replace $L$
by the centered $L_{c}$ and this then is no more positive, so that $\exp
\left( -gL_{c}\right)$ is unbounded. The point of Varadhan's theorem is to
show that this happens only on small sets so that

\begin{theorem}
(Varadhan \cite{v}) For $d=2$
\begin{equation*}
\varrho =Z^{-1}\ \exp \left( -gL_{c}\right)
\end{equation*}
with
\begin{equation*}
Z=\mathbb{E}\left( \exp \left( -gL_{c}\right) \right)
\end{equation*}
is integrable.
\end{theorem}

Varadhan defines the centered local time as the limit of Gaussian
approximations as in (\ref{3Eq2}) and uses the Chebyshev inequality to show 
that $\exp \left( -gL_{c}\right)$ is integrable for $0<g<\frac{\pi }{T}$.

A similar slightly stronger result can be obtained using Varadhan's
technique with the approximation
\begin{equation*}
L^{(2)}(\Lambda )\rightarrow L^{(2)}=L_{c}.
\end{equation*}

Fix $0<\Lambda <1$. By (\ref{eqRef}) there exists a positive
constant $k$ such that 
\begin{equation*}
L^{(2)}(\Lambda )\geq -\mathbb{E}(L(\Lambda ))\geq -k-\frac{T}{2\pi }
|\ln (\Lambda )|.
\end{equation*}
For any constant $N\geq k+\frac{T}{2\pi }|\ln (\Lambda)|$ one has 
\begin{eqnarray*}
\mathbb{P}(L_{c}\leq -N) &=&\mathbb{P}(L_{c}-L^{(2)}(\Lambda )\leq
-N-L^{(2)}(\Lambda )) \\
&\leq &\mathbb{P}\left( |L^{(2)}(\Lambda )-L_{c}|\geq N-k-\frac{T}{2\pi }%
|\ln (\Lambda )|\right) .
\end{eqnarray*}
An application of Chebyshev's inequality, using Theorem \ref{Prop} then yields 
\begin{equation*}
\mathbb{P}(L_{c}\leq -N)\leq \frac{\mathbb{E}(|L^{(2)}(\Lambda )-L_{c}|^{2})
}{\left( N-k-\frac{T}{2\pi }|\ln (\Lambda )|\right) ^{2}}\leq K\frac{\Lambda 
\ln ^{2}\Lambda}{\left( N-k-\frac{T}{2\pi }|\ln (\Lambda )|\right) ^{2}}.
\end{equation*}
In particular, for 
\begin{equation*}
\Lambda =\exp \left( -\alpha (N-k)\right),\quad 0<\alpha < \frac{2\pi }{T}
\end{equation*}
one obtains 
\begin{equation*}
\mathbb{P}(L_{c}\leq -N)\leq 
\frac{K\alpha^2}{\left(1-\frac{T}{2\pi}\alpha\right)^{2}}
\exp \left( -\alpha(N-k)\right).
\end{equation*}
Hence, $\exp \left( -gL_{c}\right) $ is integrable for $g<\frac{2\pi }{T}$.

For the Gaussian regularization an analogous result can be found in \cite{Bock}.

\section{Concluding Remarks}

\cite{FDS}, \cite{fhsw}, \cite{Hu1}, \cite{Imkeller}, \cite{Jenane}, \cite{Sandra},
\cite{NV}, in the present context of regularizations and the rate of convergence
see in particular \cite{Bock}. Yet another regularization of the
self-intersection local time is suggested by (\ref{jinky}),
namely 
\begin{equation*}
L_{\Lambda}^{(2N)}:=\sum_{\mathbf{n}: n\geq N} \langle
:\bm{\omega}^{\otimes 2\mathbf{n}}:,\varphi^{(\Lambda)}_{2\mathbf{n}}\rangle 
\end{equation*}
with kernel functions
\begin{equation*}
\varphi^{(\Lambda)}_{2\mathbf{n}}(u_{1},\ldots ,u_{2n}):=
\Theta(v-u-\Lambda) \varphi _{2\mathbf{n}}(u_{1},\ldots ,u_{2n})
\end{equation*}
where simply the range $v-u<\Lambda$ is cut out. Details of this
regularization will be discussed elsewhere.

\subsection*{Acknowledgments}

This work was financed by Portuguese national funds through FCT - 
Funda{\c c}\~ao para a Ci\^encia e Tecnologia, within the project 
PTDC/MAT-STA/1284/2012.

\end{document}